\documentstyle[epsfig]{mn}

\newif\ifAMStwofonts

\input epsf

\ifoldfss
  \ifCUPmtlplainloaded \else
    \NewTextAlphabet{textbfit} {cmbxti10} {}
    \NewTextAlphabet{textbfss} {cmssbx10} {}
    \NewMathAlphabet{mathbfit} {cmbxti10} {} 
    \NewMathAlphabet{mathbfss} {cmssbx10} {} 
  \fi
  \ifAMStwofonts
    \ifCUPmtlplainloaded \else
      \NewSymbolFont{upmath} {eurm10}
      \NewSymbolFont{AMSa} {msam10}
      \NewMathSymbol{\upi}     {0}{upmath}{19}
      \NewMathSymbol{\umu}     {0}{upmath}{16}
      \NewMathSymbol{\upartial}{0}{upmath}{40}
      \NewMathSymbol{\leqslant}{3}{AMSa}{36}
      \NewMathSymbol{\geqslant}{3}{AMSa}{3E}

    \fi
  \fi
\fi 

\ifnfssone
  \newmathalphabet{\mathit}
  \addtoversion{normal}{\mathit}{cmr}{m}{it}
  \addtoversion{bold}{\mathit}{cmr}{bx}{it}
  \newmathalphabet{\mathbfit} 
  \addtoversion{normal}{\mathbfit}{cmr}{bx}{it}
  \addtoversion{bold}{\mathbfit}{cmr}{bx}{it}
  \newmathalphabet{\mathbfss} 
  \addtoversion{normal}{\mathbfss}{cmss}{bx}{n}
  \addtoversion{bold}{\mathbfss}{cmss}{bx}{n}
  \ifAMStwofonts
    \ifCUPmtlplainloaded \else
      \UseAMStwoboldmath
      \makeatletter
      \new@mathgroup\upmath@group
      \define@mathgroup\mv@normal\upmath@group{eur}{m}{n}
      \define@mathgroup\mv@bold\upmath@group{eur}{b}{n}
      \edef\UPM{\hexnumber\upmath@group}
      \new@mathgroup\amsa@group
      \define@mathgroup\mv@normal\amsa@group{msa}{m}{n}
      \define@mathgroup\mv@bold\amsa@group{msa}{m}{n}
      \edef\AMSa{\hexnumber\amsa@group}
      \makeatother
      \mathchardef\upi="0\UPM19
      \mathchardef\umu="0\UPM16
      \mathchardef\upartial="0\UPM40
      \mathchardef\leqslant="3\AMSa36
      \mathchardef\geqslant="3\AMSa3E
    \fi
  \fi
\fi 

\ifnfsstwo
  \DeclareMathAlphabet{\mathbfit}{OT1}{cmr}{bx}{it}
  \SetMathAlphabet\mathbfit{bold}{OT1}{cmr}{bx}{it}
  \DeclareMathAlphabet{\mathbfss}{OT1}{cmss}{bx}{n}
  \SetMathAlphabet\mathbfss{bold}{OT1}{cmss}{bx}{n}
  \ifAMStwofonts
    \ifCUPmtlplainloaded \else
      \DeclareSymbolFont{UPM}{U}{eur}{m}{n}
      \SetSymbolFont{UPM}{bold}{U}{eur}{b}{n}
      \DeclareSymbolFont{AMSa}{U}{msa}{m}{n}
      \DeclareMathSymbol{\upi}{0}{UPM}{"19}
      \DeclareMathSymbol{\umu}{0}{UPM}{"16}
      \DeclareMathSymbol{\upartial}{0}{UPM}{"40}
      \DeclareMathSymbol{\leqslant}{3}{AMSa}{"36}
      \DeclareMathSymbol{\geqslant}{3}{AMSa}{"3E}
    \fi
  \fi
\fi 

\ifCUPmtlplainloaded \else
  \ifAMStwofonts \else 
    \def\upi{\pi}
    \def\umu{\mu}
    \def\upartial{\partial}
  \fi
\fi

\title{Wide-field tracking with zenith-pointing telescopes}
\author[Paul Hickson]
	{Paul Hickson \\
	The University of British Columbia, Dept. Physics \& Astronomy, 6224 Agricultural Road, 
	Vancouver, BC V6T1Z1, Canada}
\date{
      in original form 2001 February 15}

\pagerange{\pageref{firstpage}--\pageref{lastpage}}
\pubyear{2001}

\begin{document}

\maketitle

\label{firstpage}

\begin{abstract}
Equipped with a suitable optical relay system, telescopes employing low-cost
fixed primary mirrors could point and track while delivering high-quality images 
to a fixed location. Such an optical tracking system would enable liquid-mirror 
telescopes to access a large area of sky and employ infrared detectors and 
adaptive optics. Such telescopes could also form the elements of an array in 
which light is combined either incoherently or interferometrically. Tracking of 
an extended field requires correction of all aberrations including distortion, 
field curvature and tilt. A specific design is developed that allows a 10-metre 
liquid-mirror telescope to track objects for as long as 30 minutes and to point 
as far as 4 degrees from the zenith, delivering a distortion-free diffraction-limited 
image to a stationary detector, spectrograph, or interferometric beam combiner.
\end{abstract}


\section{Introduction}

The past decade has seen the introduction of large optical telescopes
employing primary mirrors that are either fixed or have restricted pointing
ability. Examples include fixed zenith-pointing liquid-mirror telescopes
(Borra et al. 1989, Hickson et al. 1994, Potter \& Mulrooney 1997,
Hickson et al. 1998) and the Hobby-Eberly Telescope (HET) which has a tilted 
primary mirror that can rotate in azimuth only. The simplification provided 
by limiting the
motion of these mirrors provides a very significant reduction in the cost
per square meter  which ranges from about one to almost two orders of 
magnitude for the case of liquid mirrors. This comes at a price of limited
pointing ability for these telescopes. The HET, which has a segmented
spherical primary mirror, employs a movable spherical aberration corrector, 
and camera or fiber feed, and can access a large area of sky and 
track objects for up to several hours. Liquid-mirrors, on the other hand, are
parabolic and have a strictly vertical axis. All astronomical liquid-mirror 
telescopes have so far operated only in a fixed zenith-pointing mode, using 
either high speed imaging (Potter \& Mulrooney 1997) or drift scanning 
detectors (Hickson et al. 1994) to compensate 
for the siderial image motion induced by the rotation of the Earth.

The low cost per unit area of liquid-mirror telescopes makes them attractive 
for surveys of faint objects. However, the zenith-pointing limitation introduces 
significant restrictions when these telescopes are used to observe faint
objects:  Infrared observations cannot easily be made because infrared arrays 
have multiplexed readouts and cannot drift-scan like CCDs. Unless the 
readout rate is very rapid, which introduces noise and data rate problems, 
the images will be degraded by the siderial drift. For drift-scanning detectors, 
the integration time is limited by the size of the detector - it is the time 
taken for an image to cross the detector, typically a minute or two. A third 
restriction is the limitation to imaging. Conventional spectroscopic observations
are not practical because of the rapid siderial image motion. The use
of adaptive optics is essentially precluded because the star images that are 
necessary for phase reference are all moving. A laser directed at the zenith
could provide a fixed reference source, but a natural guide star is still
required to provide wavefront tilt information.

A natural way to overcome these limitations would be to employ an optical
system which could direct the light received by the primary mirror to a fixed
location. This system would necessarily move as required to compensate
for the Earth's rotation for some useful period of time. Such an optical
tracking system could in principle deliver an optical beam to a stationary
detector, spectrograph, adaptive optics system or other instrument. It
could also allow the light from many individual telescopes to be brought
to a single location and combined either coherently or incoherently, as in
the proposed Large-Aperture Mirror Array (LAMA) of 18 ten-meter telescopes 
(see www.astro.ubc.ca/lmt/lama).
Clearly, an optical tracking system will be an essential component of any
multi-aperture optical/infrared interferometer that employs fixed primary mirrors.

The need for tracking is closely related to the desirability of acessing fields 
away from the zenith, thereby increasing the area of sky available to liquid mirror
telescopes. In this case, the primary consideration is
the correction of field aberrations, such as coma and astigmatism, that
occur when a parabolid is used to focus oblique rays. The problem was first 
considered by Richardson \& Morbey (1988) who found an optical design that allows 
pointing up to 7.5 degrees from the zenith, while maintaining good image quality 
over a small, but useful, instantaneous field of view. Borra (1993) concluded
that such a correction might be feasible at zenith angles as large a $45^\circ$
Wang et al. (1994), Moretto et al. (1995) and Borra et al, (1995) subsequently 
presented designs that give good image quality at large zenith angles using
actively warped mirrors.

While these studies have shown that it is possible to correct the off-axis
abberations of a parabolic primary mirror observing at high zenith angles, there is an
additional issue that has not yet been addressed. In order to image and
track an extended field of view and track, the images of celestial objects 
must not only be sharp, but the relative positions of these images within the
field must be constant. This leads to three additional requirements: First,
the image scale must not vary significantly over the full range of zenith angle.
If this were not the case, images would move radially towards or away from the
field center during the exposure causing a smearing of the images. Second,
distortion must be carefully controlled or images will again be smeared during
the exposure. If large zenith angles are involved, atmospheric dispersion and 
refraction must be continuously compensated. Third, any curvature and tilt of 
the focal plane must be eliminated or at least held constant. The optical designs
published to date do not satisfy these conditions and are therefore
not suitable for imaging an {\it extended} field of view for long exposure times.

An additional interesting development is the prospect of combining light from several
primary mirrors at a single focus. If this is done incoherently, the image brightness
increases in proportion to the total area of the primary mirrors. By combining the
light coherently (interferometrically), it is possible to achieve much higher 
resolution and sensitivity. Several prototype optical
interferometers are operational and have demonstrated phase locking and
interference at optical wavelengths (Young et al. 1998). Interferometric systems 
are being built for the Keck and VLT telescopes and other projects are planned
(von der Luehe et al. 1997, Angel et al. 1998, Booth et al. 1999). A prerequisite
for interferometry using fixed primary mirrors is that the light intercepted by each
mirror be collimated and delivered to a beam combiner, an optical system that
brings all the light to a common focus in a specific geometrical configuration.
Hence, it is of interest to consider the feasibility of tracking systems that can
deliver a collimated beam to a fixed location. 

Interferometry imposes additional requirements on any tracking system. As is well
known, the optical path lengths from the various apertures to the combined focus 
must be kept equal to within a tolerance of order $c/\delta\nu$, where $\delta\nu$
is the optical bandwidth. Furthermore, for interferometric imaging, the Abbe\' sine
condition must be preserved (the sine of the angle of incidence of all rays 
arriving at the focus must be strictly proportional to the radial position of these 
rays in the entrance pupil), and the lateral and longitudinal geometry of the exit 
and entrance pupils must match in accordance with the appropriate pupil 
magnification factors. Once a suitable design is found for the tracking system,
these additional factors can be accommodated by careful design
of the path-length equalization and beam combining systems. 

In this paper, we first analyze the restrictions that optical tracking imposes on 
the optical design. We show that this leads to a 
natural solution based on spherical symmetry. A practical design is then 
developed that allows a 10-meter liquid-mirror telescope to point and track
over an 8 degree diameter area of sky. Such a design could be suitable for 
the LAMA optical array.

\section[]{Design considerations}

It is convenient to divide the optical system into two parts: a stationary
axisymmetric system (the telescope), and a second optical system (the
relay system) which relays light from the stationary system to the final focus,
and has moving elements to provide tracking. This division is always possible 
because the primary mirror alone forms a minimal stationary axisymmetric system.
The advantage of this approach is that the two systems can be analyed and
designed individually using conventional techniques. They are then coupled to 
form the complete optical design. As we shall see, the requirements of tracking
place constraints on the the ways that the two systems can be coupled.

The telescope receives light from some angular region of the sky centred on 
the telescope axis and produces an image of this region on its focal surface. 
This region will be called the {\it accessible field of view}. The tracking system 
intercepts light from a portion of the telescope focal surface and transfers
it to a final fixed focus, where an image is formed.  The portion of the focal surface 
imaged by the relay system at any given time, and the corresponding angular region 
of the sky,  will be called the {\it instantaneous field of view}. The telescope 
focal surface may be real or virtual. In the latter case, light is intercepted by
the relay system before reaching the telescope focus.

It is not necessary that images formed at the focal surface be free from aberrations 
that affect image quality (spherical aberration, coma and astigmatism), as these
could in principle be corrected by the relay system. However, it simplifies
the design of the relay system if field-dependent aberrations such as coma and
astrigmatism are corrected by the telescope. As we shall see, tracking imposes 
additional requirements on the focal surface curvature, tilt and distortion.

The relevant geometry is illustrated in Fig.~1. Here $S$ denotes the focal surface
of the telescope. This surface is assumed to be spherical with radius $R_F$ 
(departures of the focal surface from a sphere can be treated as higher order
aberrations). $C$ denotes the centre of curvature of the focal surface and 
$V$ marks the intersection of this surface with the symmetry axis of the telescope. 
$P$ indicates the location of the telescope exit pupil. The centre of the instantaneous
field of view is denoted by $I$. The line connecting this point with $P$ corresponds 
to the path of the
principal ray (the ray that passes through the centre of the pupil). Let $\theta_P$
denote the angle between the principal ray and the axis.

\begin{figure}
  \setcounter{figure}{0}
 \epsfxsize=7cm
  \epsfbox[130 50 400 200]{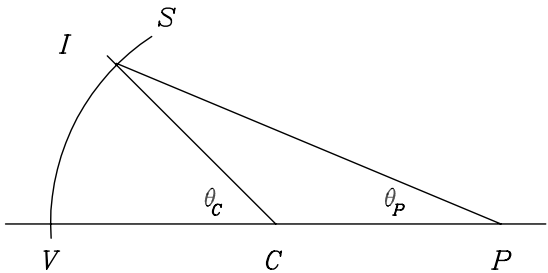}
  \caption{Geometry of the focal surface.}
\end{figure}

\subsection{Focal plane tilt}

Consider now the relay system which will deliver the light from a region centred
at $I$ to a fixed focal plane. Let $F$ denote the image of $I$ in this plane. $F$
is thus the center of the instananeous field of view in the final focal plane. 
The actual relay system may contain flat mirrors which serve only to change the
direction of the light beam. Such mirrors have no effect on the properties of the 
final image, so for the purpose of optical design and analysis it is equivalent to 
ignore these mirrors and consider only the essential elements of the system. Thus 
the beam can be considered to pass from $I$ to $F$ without any redirection by flat 
mirrors. The straight line connecting points $I$ and $F$ will be called the axis of the
relay system. For simplicity, we shall assume that the final focal plane is 
perpendicular to this axis. Generalization to a focal plane of arbitrary orientation is 
possible but will not be considered here.

Let us now suppose that the optical elements of the relay system are rotationally symmetric 
about this axis. This assumption restricts the types of relay systems that we can
consider, but it allows us to make a general analysis of the optical design problem. 
It will now be apparent that the axis of the tracking system must be normal to the
focal surface at the point $I$. Any departure from this condition will result in a
tilt of the final focal plane. Such departures are amplifed by the longitudinal magnification
factor $m_L = m_T^2$, where $m_T$ is the transverse magnification of the relay system.
Thus, {\it the axis of the relay system must pass through the centre of curvature
of the telescope focal surface}.

\subsection{Field curvature and distortion}

Curvature of the focal surface, considered alone, is not a problem as it is constant
over the accessible field of view and can therefore easily be removed by the relay system
(the necessary and sufficient condition for a flat field is that the Petzval sum of all 
optical elements in the telescope plus relay system vanish). However, tracking imposes
strict limitations on distortion, which in turn restricts field curvature, as will now be
discussed.

Distortion is a non-linear variation in image scale on the telescope focal surface.
In tracking systems, even a small amount of distortion will lead to significant differential
image motion within the instantaneous field of view. This in
turn will result in a radial elongation of images during an exposure. To avoid this,
the mapping of the sky onto the focal surface must be strictly linear. To quantify this,
let $\theta_O$ be the field angle of an object in the sky, measured with respect to the 
telescope axis, and $\phi_O$ be the azimuth angle of this object with respect to this axis,
measured from an arbitrary fiducial direction. Let $\theta_C$ be the angle between the
image of this object and the axis, at the centre of curvature of the focal surface
(see Fig.~1), and $\phi_C$ be the azimuth angle of the image. By symmetry, $\phi_C = \phi_O$.
Linearity requires that
\begin{equation}
  \theta_C = m \theta_O						\label{eq1}
\end{equation}
where $m$, the angular magnification, is a constant. Now, consider the tangential
magnification at this point. The angular displacement on the sky resulting from a small 
change $d\phi_O$ is $\sin\theta_O d\phi)$. The corresponding angular displacement 
of the image on the focal surface is $\sin\theta_C d\phi_I$. These must be related by
the same magnification factor $m$. 
\begin{equation}
  \sin \theta_C = m \sin\theta_O					\label{eq2}
\end{equation}
Clearly, equations (\ref{eq1}) and (\ref{eq2}) can only both be satisfied if $m = 1$,
which gives
\begin{equation}
  \theta_C = \theta_O						\label{eq3}
\end{equation}
Thus, tracking imposes the requirement that {\it the angle subtended at the centre of 
curvature by any point on the focal surface must be equal to the field angle of
the corresponding object}.

This leads to a condition for the radius of curvature of the focal surface, as follows.
From elementary optics, the image scale on the focal surface $ds/d\theta_0$ is 
equal to $f$, the effective focal length of the telescope. Now, we also have
$ds/d\theta_I = R$. Equation (\ref{eq2}) then requires that 
\begin{equation}
  R = f~.							\label{eq4}
\end{equation}
Thus to eliminate distortion, {\it the radius of curvature of the focal surface must equal 
the effective focal length of the telescope}.

These conditions are satisfied at the focus of a concave primary mirror, regardless of
its aspheric shape or conic constant. By symmetry, a spherical primary mirror with
radius of curvature $R_1$ has a focal surface that is is concentric with the centre of 
curvature of the primary mirror and has radius $R = R_1/2$, equal to the
primary mirror focal length. A point on this surface subtends an angle at the centre of 
curvature exactly equal to the field angle in the sky of paraxial rays that focus to this 
point. The same is true for a non-spherical primary mirror - the aspheric shape introduces
aberrations but does not change the Gaussian image parameters, which are determined
only by the curvature.  Therefore, the distortion conditions (Eqns. \ref{eq2} and \ref{eq3})) 
are exactly satistfied. To satisfy the focal-plane tilt condition (Section 2.2), a 
necessary and sufficient condition is that the axis of the tracking system be constrained 
to always pass through the 
center of curvature of the primary mirror. For a spherical primary mirror, the 
tracking system need only correct spherical aberration of the primary mirror, which
is independent of field angle, in order to achieve good image quality. For a
parabolic primary mirror, the aberrations depend on field angle and are
therefore more difficult to correct in the tracking optics. However, even in this case
it is possible to obtain excelent image quality over a wide accessible field of view, as 
will be shown in Sections 3 and 4.

\section[]{Conceptual design of tracking optics for spherical and parabolic primary mirrors}

The conceptual design of the tracking system is simplest and most easily illustrated 
for the case of a spherical primary mirror. Therefore, we begin by discussing this case.
We then extend the analysis to the case of a parabolic primary mirror, which is of
particular importance to liquid-mirror telescopes, which necessarily have a parabolic
form.

\subsection{Spherical primary mirror}

Let us suppose that the primary mirror is spherical. The tracking optics must
collimate the beam received from the primary mirror, and transmit it to a fixed 
location where it is brought to a focus. For simplicity, we shall assume that the
tracking optics deliver a fixed, aberration-free, parallel beam, which is subsequently 
refocussed by additional optical elements. Since focussing of a fixed parallel beam 
is a simple task, accomplished by a small telescope or ``beam combiner'', we do not consider
the refocussing system further, but restrict our attention to the tracking system
itself. Since this system images points on the focal surface to parallel rays, it is
just a telescope operating in reverse. Tracking is accomplished by moving this
``tracking telescope'', to follow the motion of images on the focal surface, keeping 
the axis alligned with the center of curvature. The angle that the axis of the
tracking telescope makes with the vertical is thus equal to the field angle 
(the zenith angle of the centre of the instantaneous field of view of the telescope). 
In order to keep the direction of the beam constant, a flat mirror can be placed 
at the centre of curvature and rotated by half the field angle.

\begin{figure}
 \vspace{160pt}
 \epsfxsize=7cm
  \epsfbox[40 50 440 200]{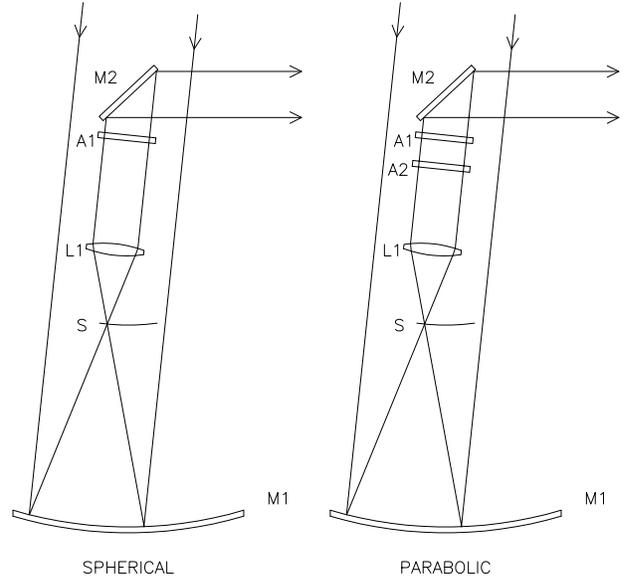}
  \caption{Conceptual tracking systems. For the spherical system shown on 
the left, light reflected from the spherical mirror M1 reaches a focus on the
spherical surface S, then is refocussed by lens L1 to produce a parallel beam
that is directed by the flat mirror M2 to a fixed location. The only aberration in this
system is spherical, which is removed by the aspheric lens A1. For a parabolic
primary, another aspheric lens A2 is placed at the image of M1 formed by
L1. It indroduces a change in optical path length equivalent to the difference
between the spherical and parabolic mirrors, making primary mirror
appear spherical.}
\end{figure}

The system is illustrated schematically in the left panel of Fig. 2, in which the tracking 
telescope is represented by a simple lens. In principle, any wide-field telescope
design that has a focal plane curvature matched to that of the focal surface
(ie $R_1/2$) can be used for the tracking telescope. However, we must also 
correct the considerable spherical abberation of the primary mirror. This might
be incorporated into the optical design of the tracking telescope, or it could be
done by adding an aspheric corrector element, centered on the axis of the
tracking telescope, as shown in the figure. Where
should such a corrector be located? In the Schmidt design, an aspheric
corrector is placed at the centre of curvature of the primary mirror, in order 
that its location does not introduce a preferred direction. In this
way, the correction that it introduces is nearly independent of field angle.
In practice, the corrector itself defines a preferred direction, which ultimately
limits the field of view of the Schmidt system. With this in mind, one might think 
that the best location for the corrector shown in Fig. 2 would be at a position 
conjugate to the centre of curvature (ie. at a real image of the centre of 
curvature). However, in our system, the corrector moves with the tracking
telescope, so there is no preferred direction on the sky no matter where we 
place the corrector along the tracking telescope axis. Therefore, the only real
consideration in its location is the optical performance of the tracking telescope 
itself. Unlike the Schmidt design, there is no limitation on the available field of view.

The above discussion has introduced the concept of tracking optics for
a fixed spherical primary mirror. Rather than present detailed optical designs, 
we now proceed to the case of a parabolic primary, which is 
particularly relevant for large lquid-mirror telescopes.

\subsection{Parabolic primary mirror}

The discussion of the proceeding section applies also to this case, but with
an additional complication. A parabolic primary mirror does not introduce
spherical aberration, but instead we have to deal with a variety of field-angle-dependent
aberrations such as coma, astigmatism and higher-order effects. The simplest
way to deal with these is to {\it optically convert the parabolic primary mirror to a 
spherical shape}. Once this is done, tracking can be accomplished exactly as
for the spherical case. We shall find, however, that the process of converting the 
parabola to a sphere introduces a preferred direction, which ultimately limits
the accessable field of view, as in the case of the Schmidt telescope.

A parabolic primary mirror can be optically converted to a sphere by placing
immediately in front of it an aspheric lens that introduces an optical path length 
difference (OPD) equal to the OPD between the parabola and a sphere of the 
same radius of curvature. For large mirrors such
a lens would be impractical, but the same effect can be obtained by locating
a similar lens at an image of the primary mirror formed by the tracking telescope.
In other words, {\it we place an aspheric corrector at a location conjugate to the
primary mirror}. To within an arbitrary additive constant, the optical thickness of 
this lens is equal to half the difference in surface height between the parabola 
and a sphere, at the corresponding point
on the primary mirror. This is illustrated schematically in the right panel of Fig 2. 

It is necessary that this aspheric corrector be exactly conjugate to the primary
mirror, both longitudinally and laterally. As the tracking telescope moves, the
position of the primary mirror with respect to the axis of the moving telescope
changes. It is therefore necessary to move the aspheric corrector, with respect
to the axis of the tracking telescope, to maintain its position and orientation 
conjugate to the primary mirror. In summary, for the case of a parabolic primary
mirror, we must add an additional aspheric corrector that is located conjugate to the
primary mirror and moves with respect to the axis of the tracking system.

\section{A practical achromatic design}

In order to see how well this concept works in practice, several trial optical designs
were investigated using CODE-V optical ray-tracing software. The simple
refractive aspheric plates described above produce good results only at the
design wavelength. Chromatic aberration from these strong aspheric surfaces
causes severe image degredation at other wavelengths. It is unlikely that 
even a colour compensated system employing two or more glasses would give
sufficient chromatic correction. Therefore, attention focussed on all-reflecting designs.
The most suitable design that was found is described in this section. The 
investigation was not exhaustive, and little attempt was made at optimization, 
so this design should be considered as illustrative only. It is possible that designs 
with even better performance might be found with a little more effort. Even so,
the results are very encouraging. This design is capable of providing sub-arcsecond
image quality at zenith angles of up to four degrees. With adaptive optics, it can
produce diffraction-limited image quality over an instantanous field of view 
limited only by atmospheric anisoplanatism. Unlike previous optical designs, it
is explicitly free from variations in distortion, focal-plane tilt and image scale
over the entire range of zenith angles. Such a system would allow exposure times
as long as 40 minutes.

The design of the tracking optics is illustrated in Fig. 3. Details of the optical system 
are given in Table 1. Light from the primary mirror focal surface enters a two-mirror 
telescope, formed by $M2$ and $M3$, that resembles a Cassegrain system. An 
aspheric mirror, $M4$, is placed at the the image of the primary mirror produced by 
the $M2 - M3$ system. A pair of relay mirrors, $M5$ and $M6$ directs the light to a second
aspheric mirror $M7$, which produces the final collimated beam. The mirror $M4$,
which converts the parabolic primary mirror to a sphere, moves laterally as the
telescope tracks in order to remain conjugate to the primary mirror. The mirror $M6$
is flat and rotates about its axis as the system tracks in order to keep the reflected
beam horzontal. The aspheric mirror $M7$, which removes the spherical aberration,
produces a horzontal collimated beam.

\begin{figure}
  \vspace{140pt}
  \epsfxsize=7cm
  \epsfbox[60 100 440 480]{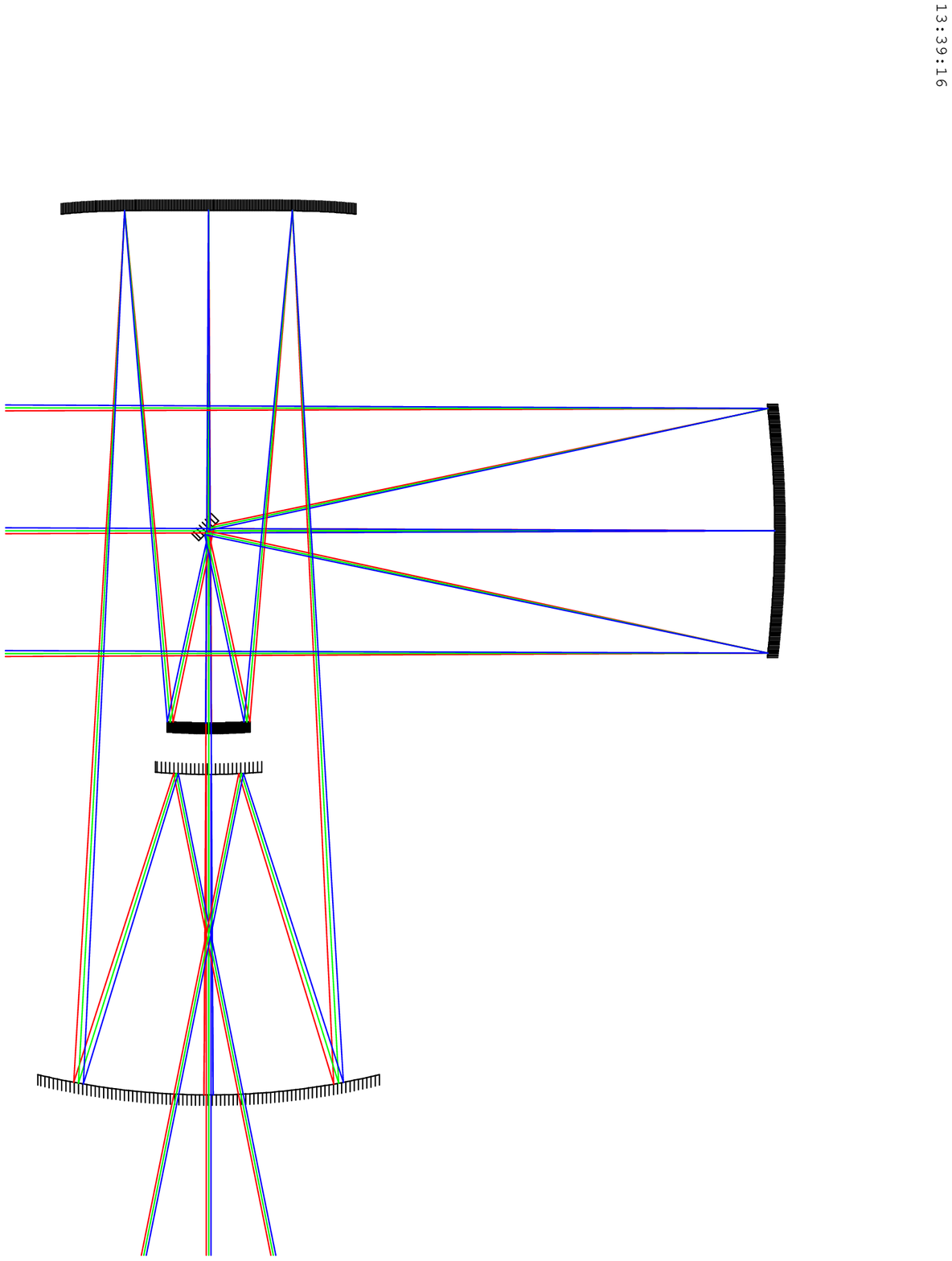}
  \caption{Parabolic tracking system at zero zenith angle. Light reflected from the
telescope primary mirror (not shown) enters the system from below. A collimated beam
is produced and directed to a fixed location where it can be brought to a focus.}
\end{figure}

\begin{figure}
  \vspace{140pt}
  \epsfxsize=7cm
  \epsfbox[60 100 460 480]{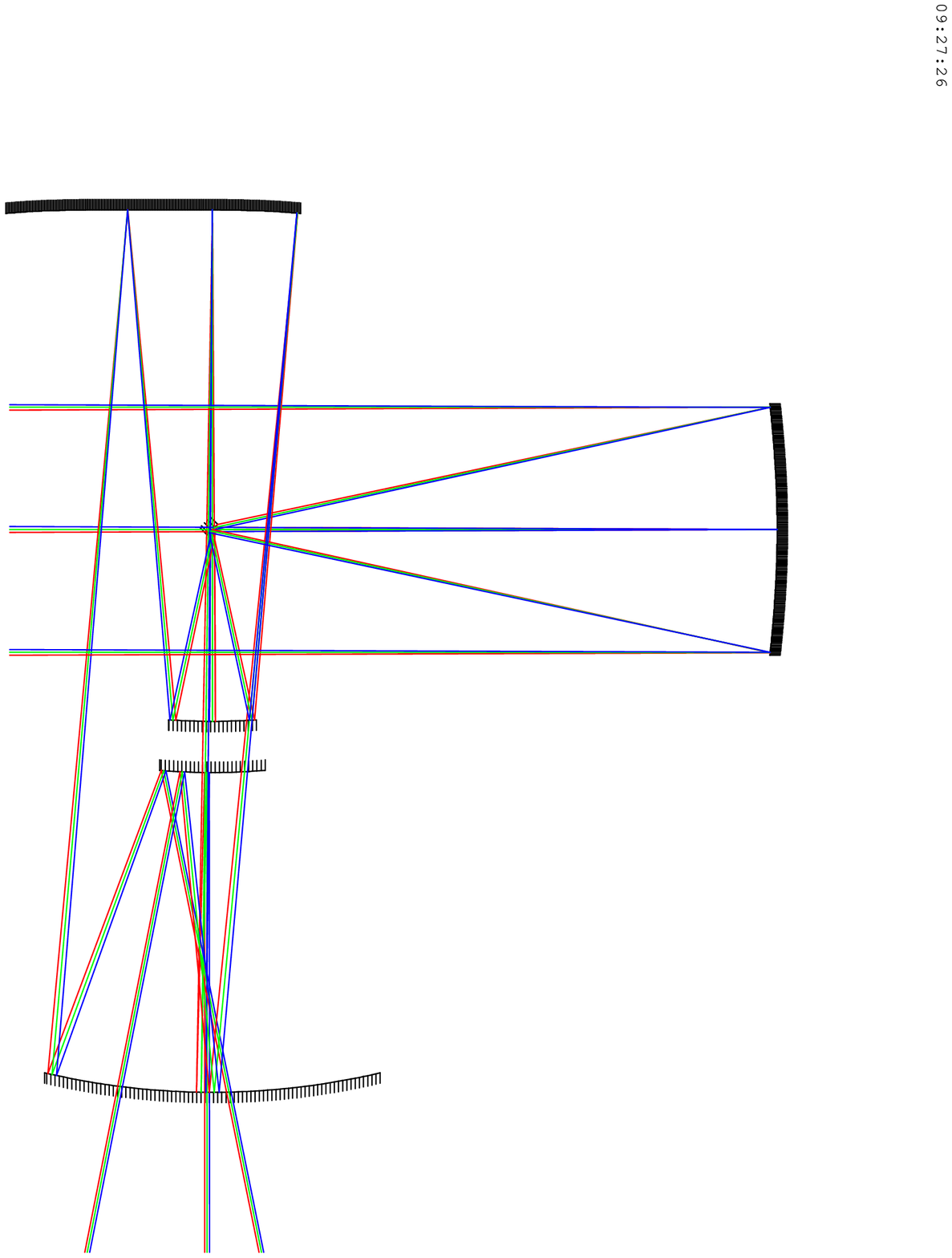}
  \caption{Parabolic tracking system at 4-degree zenith angle. The entire unit
rotates about the centre of curvature of the telescope primary mirror in order to
follow the images (this rotation is not shown in the figure). Aberrations are controlled 
by translating the aspheric mirror M4 (shown at the top of the figure), keeping it 
conjugate to the parabolic primary mirror.}
\end{figure}

\begin{table*}{\bf Table 1. Parabolic tracker optical specification}\\[1ex]
  \begin{tabular}{ccccccl}\hline\hline
Surface & radius  & axial sep. & diameter  &  conic const.  &  remarks \\
& (mm) & (mm) & (mm) \\ \hline
1 & -5000.00  & 25250.000000 & 10000  & -1.000000  & primary mirror \\ 
2 &  -960.00  & 500.000000  & 160  & -15.390077  &    \\
3 & 1140.00  & 1380.132573 & 520  & -0.402904 &      \\
4 &  $\infty$ &   800.000000  & 262 &  0.000000  & aspheric \\
& \multispan{5}{ A = 0.20217832E-08, B = -.16411329E-14,  C = 0.42870284E-19, D = -.34827625E-24\hfil} \\
& \multispan{5}{decenters: $1^\circ$: 23.221 mm, $2^\circ$: 46.509 mm, 
$3^\circ$: 69.836 mm, $4\deg$: 92.754 mm\hfil} \\

5 & -979.496000  & 300.000000 & 134 & 2.096362 &  aspheric \\
&\multispan{5}{ A = 0.64528091E-09, B = 0.28359035E-12, C = -.12423697E-15, D = 0.19904456E-19\hfil} \\    
6 & $\infty$ & 883.208421 & 48 & 0.000000 & $45\deg$ lat \\ 
7 & -1604.997119 & $\infty$ & 390 & 0.000000 & aspheric \\
& \multispan{5}{A =  0.49209044E-09, B = 0.10372729E-14, C = -.43318379E-19, D = 0.53379419E-24\hfil} \\
\hline
  \end{tabular} \\[0.5ex]
{}
\end{table*}

During tracking, $M2$, $M3$ and $M5$ rotate about the centre of curvature of 
the primary mirror, maintaining an optical axis that passes through the centre of
curvature, as shown in Fig 4. $M4$ moves with this group, but also moves
transversely with respect to the axis, in order to maintain its postion conjugate to 
the primary mirror. The flat mirror $M6$ rotates by half the angle of the 
$M2$-$M3$-$M5$ axis so that the reflected beam is horizontal. The $M6$-$M7$ 
axis remains horizontal, as does the collimated beam produced by $M7$.
The stop in this design is chosen to be at $M7$, which forms the exit pupil of 
the beam compressor. The diameter of the output beam is 0.39 m, which corresponds
to a beam compression factor, and angular magnification, of 25.6.

The performance of this system is shown by Table 2 which lists RMS spot
diameters, for three positions in the instantaneous field, as a function of 
zenith angle. The spot diameters range from 0.3 to 1.0 arcsec, over a 1
arcmin diameter field of view up to a zenith angle of four degrees. Such a
system would allow a low-latitude liquid-mirror telescope to access up to 
7\% of the sky. 

While the instantaneous field of view of of this system is quite small by the
standards of wide-field astronomy, it is quite well-suited for very-large
ground-based telescopes. Such telescopes will achieve maximum sensitivity
only by the use of adaptive optics (AO) to compensate for atmospheric seeing.
The field of view of the tracking system described here is well-matched to
the natural atmospheric isplanatic angle, which limits the instantaneous
field of view of ground-based telescopes that employ adaptive optics.

It is therefore of interest to investigate the potential of the tracking system
when adaptive optics is employed. AO systems can in principle correct telescope
aberations, but only at a given point in the field. To give good performance with
an AO system, the tracking optics design must therefore minimize {\it variations}
in aberrations across the field. To simulate an adaptive optics system, an
additional reflecting surface was added to the system described by Table 1.  This
surface was taken to be a general x-y polynomial of tenth degree, which simulates
the effect of a deformable mirror. The polynomial coefficients were optimized, at 
each zenith angle, in order to minimize the spot diameters, in the same manner
that an AO system will optimize the shape of a deformable mirror in order to
minimize the diameter of the image of a reference star. The conic and aspheric
constants of the tracking system elements were reoptimized at a zenith angle
of 2 degrees, then held constant. The specifications of this AO-optimized system
are given in Table 3, and the performance is summarized in Table 4. In addition
to the RMS spot diameters, Table 4 lists the RMS wavefront error, and the
Strehl ratio that would be achieved with a perfect AO system and no atmosphere.
The Strehl ratio (the ratio of the central intensity of the PSF to that of an unaberrated
PSF) is typically about 90\%, throughout the field, at zenith angles
up to $2^\circ$. Showing that the system is diffraction limited. At zenith
angles greater than $2^\circ$, there is a noticeable drop in Strehl ratio towards
the edge of the field. However, even at a zenith angle of $4^\circ$, the fall
in Strehl ratio is comparable to that expected due to atmospheric anisoplanatism.
This shows that over almost all of its range, the instantaneous field of
view in which near-diffraction-limited image quality can be achieved
will be limited by atmospheric anisoplanatism, rather than optical aberrations.

\begin{table}{\bf Table 2. Performance of the parabolic tracker}\\[1ex]
  \begin{tabular}{lccccc}\hline\hline\ 
Zenith Angle ($^\circ$) & 0 & 1 & 2 & 3 & 4 \\ \hline
RMS spot dia. (arcsec) \\
\hfil $-0.5$ arcmin		& 0.62	& 0.73	& 0.79	& 1.03	& 1.03 \\
\hfil $~0.0$ arcmin		& 0.30	& 0.31	& 0.36	& 0.68	& 0.69 \\
\hfil $+0.5$ arcmin	 	& 0.62	& 0.42	& 0.36	& 0.79	& 0.96 \\
\hline
\end{tabular} \\[0.5ex]
{}
\end{table}

\begin{table*}{\bf Table 3. AO-optimized tracker optical specification}\\[1ex]
  \begin{tabular}{ccccccl}\hline\hline
Surface & radius  & axial separation & diameter  &  conic constant   &  remarks \\
& (mm) & (mm) & (mm) \\ \hline
1 & -5000.00  & 25250.000000 & 10000  & -1.000000  & primary mirror \\ 
2 &  -960.00  & 500.000000  & 160  & -1.253120  &    \\
3 & 1140.00  & 1380.132573 & 520  & -0.004374 &      \\
4 &  $\infty$ &   800.000000  & 262 &  0.000000  & aspheric \\
& \multispan{5}{ A = 0.19630681E-08, B = 0.35801621E-15,  C = 0.50652415E-19, 
D = -.14431399E-24 \hfil} \\
& \multispan{5}{decenters: $1^\circ$: 24.200 mm, $2^\circ$: 48.052 mm, $3^\circ$: 71.166 mm, 
$4^\circ$: 93.303 mm \hfil} \\

5 & -979.496000  & 300.000000 & 134 & 4.310945 & aspheric \\
& \multispan{5}{ A = 0.65349064E-09, B = 0.36582659E-14, C = 0.19443600E-17, 
D = -.16005167E-21 \hfil} \\    
6 & $\infty$ & 883.208421 & 48 & 0.000000 & $45\deg$ tilt \\ 
7 & -1604.997119 & 258.000000 & 390 & 0.000000 & aspheric \\
& \multispan{5}{ A =  0.46139661E-09, B = 0.82228387E-15, C = -.13956758E-19, 
D = 0.19158183E-24 \hfil} \\
8 & $\infty$ & $\infty$ & 385 & 0.000000 & XY-polynomial \\
\hline
  \end{tabular} \\[0.5ex]
{}
\end{table*}

\begin{table}{\bf Table 4. Performance of the AO-optimzed tracker}\\[1ex]
  \begin{tabular}{lccccc}\hline\hline\ 
Zenith Angle ($\deg$) & 0 & 1 & 2 & 3 & 4 \\ \hline
RMS spot dia. (mas) \\
\hfil $-0.5$ arcmin		& 16.3	& 13.3	& 10.8	& 23.9	& 43.4 \\
\hfil $~0.0$ arcmin		& 13.2	& 13.9	& 16.9	& 22.9	& 30.8 \\
\hfil $+0.5$ arcmin	 	& 16.3	& 15.8	& 17.2	& 28.4	& 43.5 \\
RMS OPD (nm) \\
\hfil $-0.5$ arcmin		& 110	& 100	& 123	& 228	&  335 \\
\hfil $~0.0$ arcmin		& 73	& 79	& 104	& 149	&  147 \\
\hfil $+0.5$ arcmin	 	& 110	& 99	& 75	& 198	&  334 \\
Strehl Ratio ($\lambda = 2$ um)\\
\hfil $-0.5$ arcmin		& 0.89	& 0.91	& 0.86	& 0.60	&  0.33 \\
\hfil $~0.0$ arcmin		& 0.95	& 0.94	& 0.90	& 0.80	&  0.80 \\
\hfil $+0.5$ arcmin	 	& 0.89	& 0.91	& 0.95	& 0.68	&  0.33 \\
\hline
\end{tabular} \\[0.5ex]
{}
\end{table}

\section[]{Conclusions}

We have shown in this paper that a telescope with a fixed primary mirror can
track celestial objects for an extended period of time. But, to accomplish
this requires more than just a corrector system that produces sharp images.
There must also be no variations in the relative positions of these 
images in the field, throughout the tracking range. This requirement places
additional constraints on the optical system that have been outlined in
Section 2. We have established
these conditions can be satisfied by a system of moving optics (the tracking
system), providing that tracking is accomplished by rotating this system
about the center of curvature of the fixed primary mirror. The tracking
system is simplest in the case of a spherical primary mirror, where pointing
and tracking over very large zenith angles is possible, although in practice 
this would be limited by the effects of atmospheric refraction and dispersion.
A parabolic primary mirror can also be accommodated, by means of a 
conjugate aspheric corrector, although the range of zenith angle is more
restricted in this case.

The tracking system may either focus the light directly on a detector,
which moves with the system, or redirect it to a fixed location. The latter
case is useful to feed a large or mechanically sensitive instrument, such
as a high-resolution spectrograph, and for the combination of light from
an array of mirrors.

Telecopes using fixed primary mirrors can be built at much less cost than
those that are fully steerable. Liquid-mirror telescopes that have so far
been built have typical costs about 5\% of those of conventional telescopes
of similar aperture. Until now, such telescopes have operated only in a
zenith-pointing mode, using time-delay integration CCD detectors to
compensate for siderial motion. This has restricted them to short exposure
times, typically less than two minutes duration, and allowed them to access
only a small fraction (less than 0.5\%) of the sky. Tracking optics would
increase the available sky and available exposure times by an order of
magnitude.

Large ground-based telescopes can achieve maximum sensitivity for
point-like sources only if they are near diffraction limited (the well-known
$D^4$ advantage). Adaptive optics is essential to this. At present,
atmospheric isoplanatism restricts the field of view to less than an arcmin.
Future multi-conjugate AO systems should increase this limit, but not
likely by an order of magnitude. For this reason, we have focussed on optical
designs that are matched to this atmospheric limit. As an example, a particular
design was presented that is capable of achieving diffraction-limited
image quality in the near-infrared at zenith angles up to four degrees.
Such a system would allow very large liquid-mirror telescopes to achieve
diffraction-limited resolution and sensitivity, and to track fields for up to 30 min. 
By providing a collimated beam to a fixed location, it
would allow light from many primary mirrors to be combined, either
incoherently or coherently, or to be directed to large instruments.

The optical designs presented here are intended to be illustrative only. We have
not extensively explored the parameter space of these designs, and further
optimization is possible. More analysis is required to explore the ultimate 
limits of such systems. In particular, by specifically employing a wide-field 
telescope design in the tracking system, it may be possible to significantly 
increase the instantaneous field of view.

\section*{Acknowledgments}

This work was supported by grants from the Natural Sciences and
Engineering Research Council of Canada.

\label{lastpage}

\end{document}